**Probing weakly hybridized magnetic molecules by spin-polarized tunneling**

*Emil Sierda[1,2,\*], Micha Elsebach[1], Roland Wiesendanger[1] and Maciej Bazarnik[1,2]*

[1]*Dept. of Physics, University of Hamburg, Jungiusstrasse 11, D-20355 Hamburg, Germany*

[2]*Institute of Physics, Poznan University of Technology, Piotrowo 3, 60-965 Poznan, Poland*

\* *esierda@physnet.uni-hamburg.de*

**Advances in molecular spintronics rely on the in-depth characterization of the molecular building blocks in terms of their electronic and, more importantly, magnetic properties. For this purpose, inert substrates that interact only weakly with adsorbed molecules are required in order to preserve their electronic states. Here, we investigate the magnetic-field response of a single paramagnetic 5,5'-dibromosalophenatocobalt(II) (CoSal) molecule adsorbed on a weakly interacting magnetic substrate, namely Fe-intercalated graphene (GR/Fe) grown on Ir(111), by using spin-polarized scanning tunneling microscopy and spectroscopy (SP-STM/STS). We have obtained local magnetization curves, spin-dependent tunneling spectra, and spatial maps of magnetic asymmetry for a single CoSal molecule, revealing its magnetic properties and coupling to the local environment. The distinct magnetic behavior of the Co-metal center is found to rely strictly on its position relative to the GR/Fe moiré structure, which determines the level of hybridization between the GR/Fe surface $\pi$-system, the molecular ligand $\pi$-orbitals and the molecular Co-ion $d$-orbital.**

Molecular-based systems are promising candidates for nanospintronic devices, the main examples being single molecular magnets (SMMs) [Bogani, 2008, Schwöbel, 2012, Dreiser, 2016, Gragnaniello, 2017, Paschke, 2019], phthalocyanines [Iacovita, 2008, Javaid, 2010, Atodiresei, 2010, Brede, 2010, 2012, Mugarza, 2011, Avvisati, 2018, 2018, Czap, 2019], and carbon-based magnetic heterostructures [Gambardella, 2009, Kawahara, 2012, Brede, 2014]. In most of these studies the molecules' orbitals were strongly hybridized with the substrates' electronic states [Schwöbel, 2012,

Iacovita, 2008, Javaid, 2010, Atodiresei, 2010, Brede, 2010, 2012, Mugarza, 2011, Gambardella, 2009, Kawahara, 2012, Brede, 2014, Czap, 2019]. Based on this fact, there have been concerns about how to separate contributions of the molecules and the substrate to the measured magnetic signal. Recent experiments performed with molecules adsorbed on inert graphene (GR) [Dreiser, 2016, Gragnaniello, 2017, Paschke, 2019] have attempted to reduce the effect of hybridization on the magnetism of the molecule. However, these studies involved either spatially averaging techniques [Dreiser, 2016, Gragnaniello, 2017] or made use of inelastic spin excitations [Paschke, 2019] which do not address the spatial distribution of spin-polarized states of single adsorbed molecules. At the same time, spatially averaging techniques indicate that a Co-intercalated graphene (GR/Co) substrate exhibits significant hybridization with adsorbed molecules [Avvisati, 2018, 2018]. However, spatial variations of the electronic properties of intercalated graphene, due to its moiré structure [Decker, 2014], can result in a locally low level of substrate-molecule hybridization [Bazarnik, 2013]. In parallel, there have been a number of experimental as well as theoretical studies on the possibility to create planar and all-spin logic devices (SLD) using salophene-based molecules [DiLullo, 2012, Bazarnik, 2016, Sierda, 2017, García-Fernández, 2017]. Here we take another step towards to this ultimate goal and demonstrate the detection and manipulation of the direction of the magnetic moment ($\mu_m$) of individual CoSal molecules, based on the low level of hybridization with a GR/Fe substrate. The substrate stabilizes $\mu_m$ and provides a stable magnetic reference required for SP-STM/STS studies [Wiesendanger, 2009].

For the following investigations, we have intentionally prepared a spatially heterogeneous sample system consisting of areas of pristine GR/Ir(111) as well as areas of GR/Fe. The subsequently deposited CoSal molecules were found only on the latter regions and prefer to adsorb along step edges, followed by step-flow growth leading to molecular assemblies located near the step edges. The moiré structure of the GR/Fe substrate [Decker, 2014] leads to an additional preferential adsorption of the CoSal molecules away from the moiré's top sites. This behavior can be explained by the combined action of intermolecular Van-der-Waals forces and repulsive dipolar interactions between molecules and the substrate which drive the molecules away from the moiré's top sites and influence their arrangement relative to one another [Bazarnik, 2013]. An example for such a molecular assembly consisting of 27 individual molecules is presented in Fig. 1a. In this STM topography image, starting from the left, we can observe an area of GR/Fe, an area of CoSal/GR/Fe and, separated by an upward step-edge, an area of GR/Ir(111). A schematic cross-section of the sample structure is provided in Fig. 1b. Its location within the STM image of Fig. 1a is marked by a white dashed line. At a sample bias voltage of $U = +450$ mV (as used for recording the data in Fig. 1a) electronic states located on the Co-metal centers of the molecules are dominant. Structural models of the molecules and the positions of the moiré top sites are superimposed onto the STM data in the lower part of Fig. 1a.

Next, we focused on an area around a single molecule, as marked by a white square in Fig. 1a. Fig. 1c and Fig. 1d show topographic STM images of that region obtained with two different sample bias voltages: $U = +50$ mV and $U = +450$ mV, respectively. In order to enhance the spatial resolution, a molecule-terminated probe tip was used to obtain those images. By tunneling at a low bias of $U = +50$ mV (within the CoSal HOMO-LUMO gap) with such a probe tip, we can obtain intramolecular resolution of the CoSal molecule [Repp, 2005]. One can easily identify the protrusion in the middle of the molecule originating from the center Co atom, two protrusions arising from the Br end atoms and the molecular backbone formed by the rest of the atoms. A schematic drawing of that area is shown in Fig. 1e, where ball-and-stick models of the center molecule as well as of other surrounding molecules are depicted on top of a model of the GR/Fe substrate's atomic structure (red and black lines representing the GR and Fe lattices, respectively). The particular molecule we are focusing on in the following is adsorbed in between the hcp and fcc areas of the GR/Fe moiré structure. This area appears slightly protruded, creating a saddle-type line from one top-site of the moiré to another. The CoSal molecule of our focus is adsorbed slightly off the saddle line towards the hcp region. It is of central importance for the discussion later that this position exhibits a lower spin polarization of the electronic states than neighboring hcp and fcc regions [Decker, 2014].

Based on the bias-dependent differential tunneling conductance ($dI/dU$) maps we concluded that the maximum signal at the Co-centers for most of the molecules within the assembly is observed at $U =$

+450 mV. Therefore, we selected this particular bias to map the spin-resolved *dI/dU* signal as a function of an externally applied magnetic field ($B = 0.75$ T $\rightarrow$ 5.25 T $\rightarrow$ -5.25 T $\rightarrow$ 0.75 T) by using a magnetic SP-STM probe tip [Wiesendanger, 2009]. Based on that data we have obtained the spin-resolved differential tunneling conductance signal (*G*) as a function of magnetic field (details in supplementary information). It is spatially averaged over two areas; one including the position of the molecule's Co atom and the other covering an hcp area of GR/Fe. The resulting locally measured magnetization curves [Meier, 2008] are shown in Fig. 2a. For the interpretation of the measured data, it is important to note that the SP-STM probe tip used for these experiments is magnetically soft and requires only $B = 0.2$ T in order to fully align its magnetization direction with $\vec{B}$. Therefore, while crossing $B = 0$ T one will always observe a change in the spin-resolved *G*-signal intensity due to a change of the SP-STM probe tip's magnetization direction. The measured values of *G* obtained for GR/Fe as a function of $\vec{B}$-field reveal a characteristic magnetic hysteresis. There are two changes of the substrate's magnetization direction occurring: after increasing the $\vec{B}$-field from 4.5 T to 5.25 T and decreasing from -4.5 T to -5.25 T. This behavior is expected for these $\vec{B}$-field values as reported by Decker *et al.* [Decker, 2014]. The GR/Fe loop is inverted here due to the different sample bias voltage used to obtain the data. The deduced GR/Fe magnetization directions are ↓ for forward [-5.25 T, 4.5 T] and ↑ for backward [-4.5 T, 5.25 T] $\vec{B}$-field sweeps (see Fig. 2a). In contrast, the CoSal response to the $\vec{B}$-field variation is markedly different. It is still mirror-symmetric with respect to $B = 0$ T and is affected by the above-mentioned changes of the GR/Fe magnetization direction. However, we can observe a very pronounced increase in *G*-signal as the $\vec{B}$-field rises from 1.5 T to 3.75 T. This suggests that the z-component of the magnetic moment $\mu_z$ of the molecule's Co-center is aligning with the $\vec{B}$-field. Schematic drawings indicating the magnetization directions of all parts of the magnetic tunnel junction for $B = 0.75$ T, 3.75 T, and 5.25 T are provided in Fig. 2b. Upon reaching the value of the $\vec{B}$-field for which the GR/Fe magnetization direction changes, the magnetic moment of the CoSal molecule aligns fully with the direction of the substrate's magnetization and the external magnetic field. A drop in *G*-signal intensity is strictly connected to changes of the substrate's magnetization. Subsequently, the external magnetic field is lowered and no significant changes are observed until $B = 0$ T. Upon the SP-STM tip's magnetization reversal the behavior described above is repeated for the opposite $\vec{B}$-field direction.

Two spin configurations as outlined in Fig. 2b (for $B = 0.75$ T and 3.75 T) have been used for further SP-STS experiments in order to visualize the spin-dependent local density of states (LDOS) distributions and the differences between them. Before each SP-STS curve has been recorded, we stabilized the SP-STM probe tip above a non-magnetic part of the substrate, *i.e.* bare GR/Ir(111) (without the Fe-intercalation layer), in order to guarantee a constant and spin-polarization independent sample – tip separation [Kubetzka, 2003]. The SP-STS data presented in Fig. 3a was measured on the Co-center of the CoSal molecule depicted in Fig. 1c at $B$-field values of 1.5 T and 4.5 T. The changes in the measured spectra reflect the behavior observed for the magnetization curves in Fig. 2a, and one can clearly distinguish three bias regions for which spin-dependent tunneling effects are most pronounced: one around $U = -250$ mV, one around the peak at $U = +500$ mV, and another around $U = +1$ V. It is important to note that at the same time, the SP-STS data on bare GR/Ir(111) (see Fig. 3b) does not show any changes when comparing the curves acquired at different $\vec{B}$-field values. Hence, one can conclude that the changes observed for the CoSal molecule originate purely from its response to the external $\vec{B}$-field. The change in the spectra of CoSal does not induce any change in the spectra of GR/Fe. A similar experiment in which the magnetization direction of GR/Fe has been inverted is presented in the supplementary information (Fig S1). There, one can clearly see a change in the SP-STS data of GR/Fe and how that influences the behavior of CoSal. After revealing the energetic positions of the spin-polarized states, we map their spatial distributions by spin-resolved $dI/dU$ maps in Fig. 3c-e (HOMO at $U = -250$ mV, LUMO at $U = +450$ mV, and LUMO+1 at $U = +950$ mV, respectively) and their corresponding spin asymmetry in Fig. 3f-h. The spin asymmetry distribution for $U = -250$ mV is close to zero, for $U = +450$ mV it is of positive sign and particularly strong over the molecules' center atoms, while for $U = +950$ mV it is of negative sign and localized in the same area.

The single-molecule magnetization curves recorded in the present study and displayed in Fig. 2a are only part of a full hysteresis loop, *i.e.* magnetic saturation is not reached, which is caused by two factors. On the one hand, the GR/Fe magnetization direction changes in relatively low $\vec{B}$-fields and CoSal follows this behavior. On the other hand, using a Brillouin function we have estimated that the $\vec{B}$-field needed to fully align the molecule's magnetic moment (assuming it preserves spin ½) at our measurement temperature of 6.5 K would be as high as 29 T. In order to determine the full magnetization curve, the measurements would have to be performed at temperatures of ~1 K or lower for the experimentally accessible $\vec{B}$-field (up to 6 T). At low $\vec{B}$-field the molecule's magnetic moment is stabilized by superexchange interactions with the substrate. The same explanation has also been proposed for the similar system of phthalocyanine molecules adsorbed on a GR/Co substrate [Avvisati, 2018, 2018]. The spatially averaging techniques as used by Avvisati *et al.* suggested either ferro- or antiferromagnetic interactions of the molecule with the substrate, depending on the type of metal centers of different molecules. However, by using local probe techniques, we show here that the different adsorption sites are non-equivalent and exhibit different degrees of magnetic interaction strengths between molecule and substrate. Hence, it is possible to manipulate the molecule's magnetic moment by an external $\vec{B}$-field. Based on previously reported DFT calculations for CoSal molecules we expect the $d_{xz}$-orbital to carry a non-paired electron [DiLullo, 2012, Bazarnik, 2016]. Its geometry can lead to both ferro- and antiferromagnetic superexchange coupling between the GR/Fe substrate and the CoSal molecule. The latter is observed in our experiment. Therefore, its path is as follows: GR C π → CoSal O π + N π → Co $d_{xz}$. The strength of this interaction can vary depending on the number of CoSal π-orbitals effectively interacting with the GR π-system. That is the reason why a sufficiently strong $\vec{B}$-field can overcome this interaction and act on the molecule's magnetic moment in some of the adsorption configurations, including the one presented here. Following the equipartition theorem, the average thermal kinetic energy for this system is $E_{k(z)} = ½k_BT = 0.28$ meV. The difference in the Zeeman energy for $B = 1.5$ T and 4.5 T, *i.e.* magnetic fields for which the SP-STS data of Fig. 3a has been obtained, is $\Delta E_Z = 0.3$ meV. As $\Delta E_Z$ is comparable to $E_{k(z)}$, no spin splitting is expected to be observed and Fig. 3a shows that indeed no spin splitting is visible. The spin asymmetry observed in Fig. 3c-e is low because, as discussed above, we only partly align the magnetic moment of the CoSal molecule with the $\vec{B}$-field. The effect is small and affecting only one direction (normal to the substrate). In Fig. 3f an almost vanishing spin asymmetry is observed, caused by a low spin polarization of the electronic states at that particular bias voltage. The signs of the spin asymmetries observed in Fig. 3g and Fig. 3h are in agreement with the SP-STS differences as revealed in Fig. 3a. They indicate that the two spin-polarized molecular orbitals located on the N, O and Co atoms of CoSal exhibit two directions with respect to GR/Fe: antiparallel originating from Co d orbitals and parallel originating from the O π and N π orbitals.

In conclusion, we have observed spin-polarized tunneling to molecular orbitals of a paramagnetic molecule adsorbed on a ferromagnetic GR/Fe substrate. The distinct adsorption site and the effectively low hybridization resulted in relatively weak magnetic coupling of CoSal with the GR/Fe enabling us to address the spin-dependent behavior of the molecule independent from the substrate. Moreover, we mapped the spin polarization distribution of two molecular orbitals by bias-dependent SP-STS experiments. Being able to manipulate and read-out the magnetic state of individual adsorbed molecules independent from the substrate's magnetic state will be of great importance for the design and realization of molecular spintronic devices.

**Methods**

All experiments were performed in a UHV system equipped with a low temperature SP-STM and two preparation chambers: one for substrate cleaning and CVD growth and another for molecule as well as metal deposition [Wittneven, 1997]. Electrochemically etched tungsten tips cleaned by standard *in situ* procedures and coated with ~50 ML of Fe were used as probes for our SP-STM studies. The Ir(111) single crystals were cleaned by repeated cycles of Ar$^+$ sputtering (800 V, 5E-6 mbar), annealing at temperatures ranging from 900 K to 1500 K in the presence of O$_2$ and a flash annealing at ~1500 K. The graphene layer was grown *in situ* on Ir(111) by decomposition of ethylene molecules following the procedure described in ref. [N'Diaye, 2008]. Nearly full layer intercalation of Fe was achieved *in situ* following the procedure of ref. [Bazarnik, 2015]. Molecules were deposited from thermally cleaned aluminum nitride (AlN) crucibles by thermal sublimation under UHV conditions directly onto the substrate held at room temperature. Samples after preparation were transferred *in vacuo* to the SP-STM setup and cooled down to the measurement temperature of 6.5 K. The exact positions of the molecules within the assembly presented in Fig. 1a have been determined by a combination of energy-dependent *dI/dU* maps revealing the electronic states of different functional groups within the molecules and high-resolution STM images obtained with a molecule terminated probe tip. Lattice constants for Gr and Fe lattices used in Fig. 1e have been extracted from ref. [Zeller, 2014]. A lock-in detection technique was used to obtain *dI/dU* maps and point spectroscopy data. The *dI/dU* maps were recorded simultaneously with the STM topography in the constant-current mode. The spin asymmetry is defined here as $dI/dU_{asym}(U) = dI/dU\uparrow\uparrow(U) - dI/dU\uparrow\downarrow(U) / dI/dU\uparrow\uparrow(U) + dI/dU\uparrow\downarrow(U)$ where the directions ↑ and ↓ refer to the magnetization of the SP-STM probe tip and the sample. The *G* values have been averaged over areas of ~0.1 nm$^2$ and ~0.7 nm$^2$ for the molecule's Co centers and the GR/Fe hcp regions respectively. All data has been processed using MATLAB and Gwyddion [Nečas, 2012] software.

Tunneling parameters:

Fig. 1a: $U$ = +450 mV, $I$ = 55pA;

Fig. 1c: $U$ = +50 mV, $I$ = 50pA;

Fig. 1d: $U$ = +450 mV, $I$ = 50pA;

Fig. 2: Data is extracted from *dI/dU* maps recorded with: $U$ = +450 mV, $I$ = 55pA, $f_{mod}$ = 971 Hz, $V_{mod}$ = 50 mV$_{rms}$;

Fig. 3a: $U_{stab}$ = +1 V, $I_{stab}$ = 55pA, $z_{off}$ = 100 pm, $f_{mod}$ = 971 Hz, $V_{mod}$ = 50 mV$_{rms}$, every line represents an average over 5 individual spectra;

Fig. 3b: $U_{stab}$ = +1 V, $I_{stab}$ = 200pA, $z_{off}$ = 0 pm, $f_{mod}$ = 971 Hz, $V_{mod}$ = 50 mV$_{rms}$, every line represents an average over 5 individual spectra;

Fig. 3c: $U$ = -250 mV, $I$ = 55pA, $f_{mod}$ = 971 Hz, $V_{mod}$ = 50 mV$_{rms}$;

Fig. 3d: $U = +450$ mV, $I = 55$ pA, $f_{mod} = 971$ Hz, $V_{mod} = 50$ mV$_{rms}$;

Fig. 3e: $U = +950$ mV, $I = 55$ pA, $f_{mod} = 971$ Hz, $V_{mod} = 50$ mV$_{rms}$.

**Acknowledgements**

We gratefully acknowledge financial support from the Office of Naval Research via grant No. N00014-16-1-2900. M.B. additionally acknowledges the support of the National Science Centre, Poland under grant nr. 2017/26/E/ST3/00140. We are grateful to J. Wiebe for insightful discussions and to B. Bugenhagen for providing us with molecules and insightful discussions.

**Author contributions**

M.B. conceived the experiment. E.S. performed the measurements and analyzed the data. M.E. supported the measurements. R.W. and M.B. supervised the work. E.S., R.W., and M.B. wrote the manuscript. All authors discussed the results and commented on the manuscript.

**Competing interests**

The authors declare no competing financial interests.

**Figures**

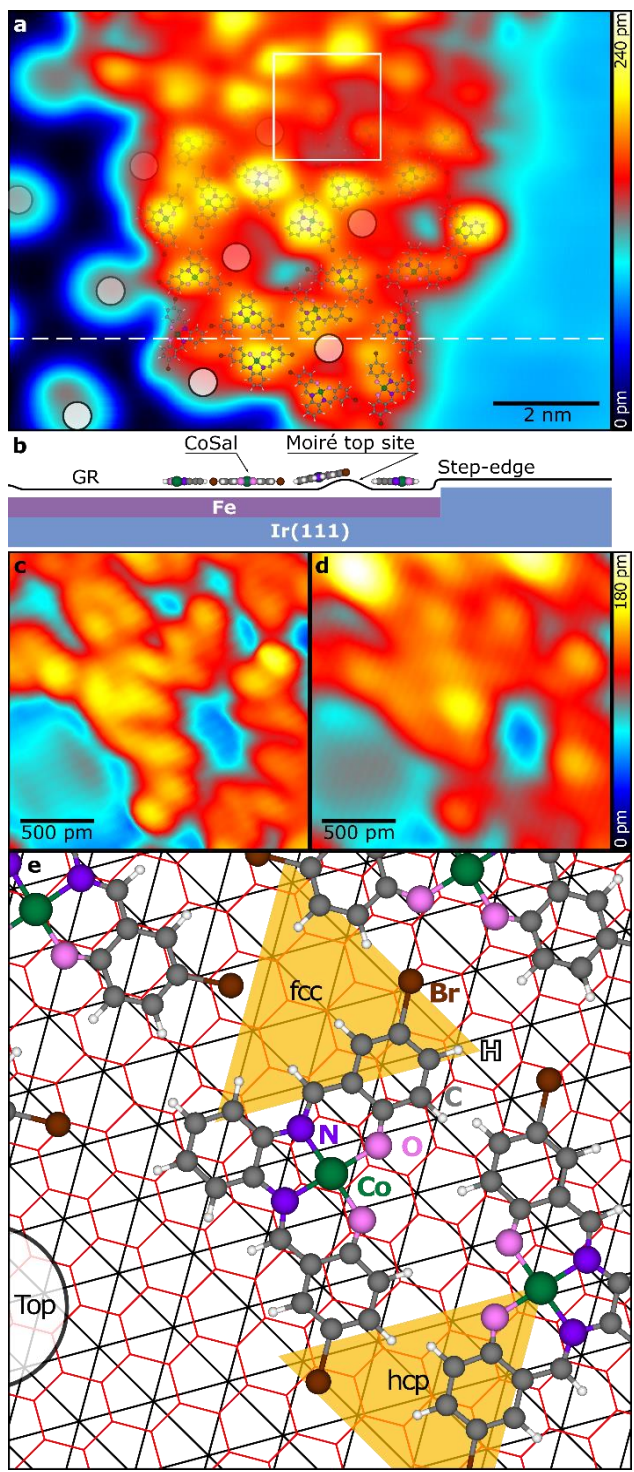

**Figure 1:** STM topographs of CoSal molecules adsorbed on GR/Fe and corresponding structure model of the imaged area. (a) STM overview image of the molecular assembly used for subsequent higher-resolution measurements ($U = +450$ mV, $I = 55$ pA). Ball-and-stick models of the molecules within the assembly are superimposed on the measured STM data, while the positions of the circles represent top sites of the moiré pattern originating from the intercalated GR layer. (b) A schematic

cross-section along the white dashed line in the STM image shown in (a). (c)-(d) Bias-dependent STM images of a single CoSal molecule within the assembly obtained with a molecular probe tip in the area marked with a white square in (a): $U = +50$ mV for (c) and $U = +450$ mV for (d), while $I = 50$ pA. (e) Ball-and-stick models of the molecules overlaid on the atomic lattice of the GR/Fe substrate (GR: red, Fe: black) as deduced from (a). The individual atomic species within the CoSal molecule as well as the specific regions of the GR/Fe moiré structure are labelled.

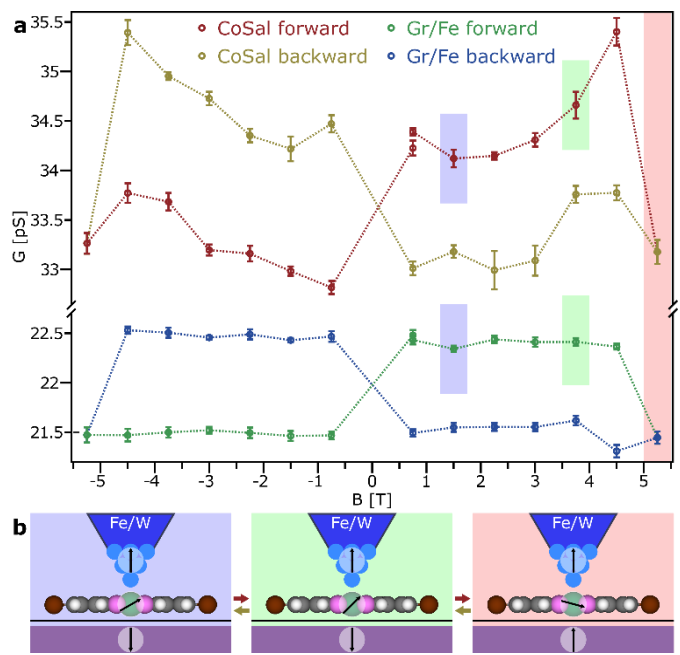

**Figure 2:** Magnetic response of a single adsorbed CoSal molecule as imaged in Fig. 1c and the bare GR/Fe substrate to an external out-of-plane magnetic field $\vec{B}$. (a) Spin-resolved differential tunneling conductance $G$ extracted from the measured $dI/dU$ signal averaged over the molecule's Co-metal center and over a bare GR/Fe area next to the molecular assembly. The error bars correspond to the standard deviation of the spatially averaged $dI/dU$ signals. (b) Schematic drawings indicating the magnetization directions of the Fe-coated W-probe tip, the Co-center of the CoSal molecule and the GR/Fe layer for three different $B$-field values as color coded in (a): 1.50 T (blue), 3.75 T (green), and 5.25 T (red).

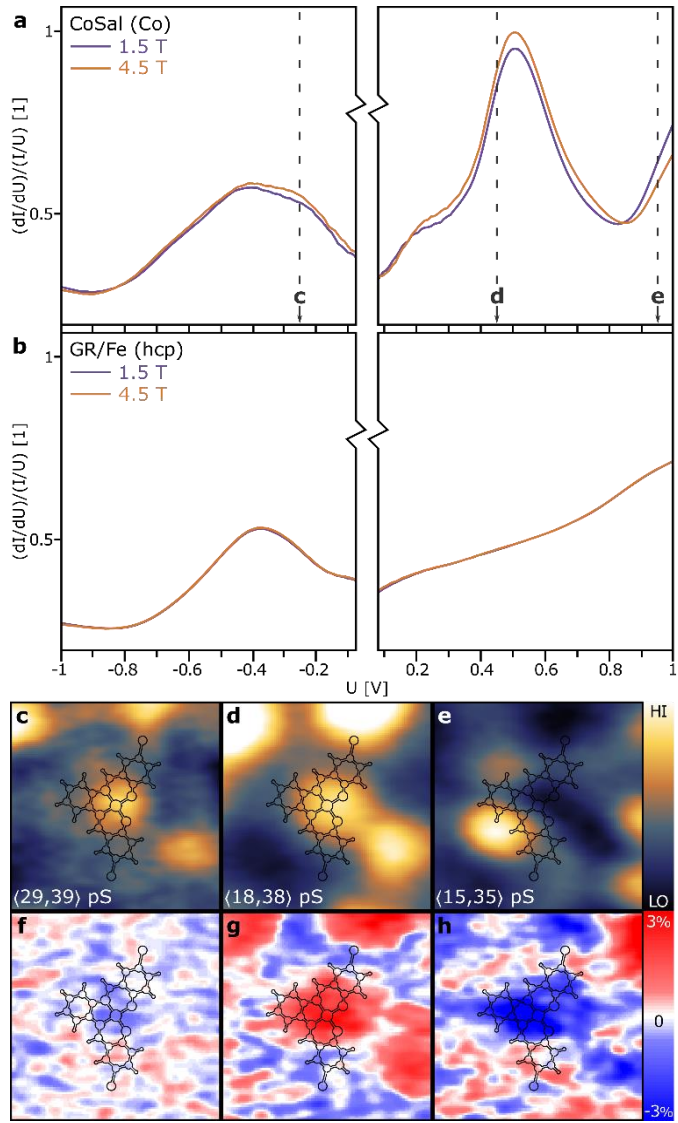

**Figure 3:** Local SP-STS data and SP-STM images of a single CoSal-molecule measured in different external *B*-fields. (a)-(b) Normalized SP-STS obtained on the CoSal molecule's Co-center (a) and on a bare GR/Fe area next to the molecular assembly (b) at different magnetic field strengths: *B* = 1.5 T and 4.5 T. The measured spin-resolved differential tunneling conductance *dI/dU* has been divided by *I/U* and plotted as a function of bias voltage *U*. (c)-(e) Spatially resolved *dI/dU* maps for three different bias voltages: *U* = -250 mV (c), +450 mV (d), and +950 mV (e) as marked in (a) by dashed lines (*I* = 55 pA). (f)-(h) Spin asymmetry maps for the bias voltages as presented above in (c)-(e). All maps in (c)-(h) have been superimposed by the structural model of the CoSal molecule.